\documentclass{WileyMSP-template}

\usepackage[square, numbers, comma, sort&compress,super]{natbib}
\usepackage{ragged2e}%

\usepackage[font={sf,bf},textfont=md]{caption}

\usepackage[colorlinks,
linkcolor=blue,
anchorcolor=blue,
citecolor=blue,
]{hyperref}

\usepackage{amsmath}
\usepackage{amsfonts}
\usepackage{amssymb}
\usepackage{graphicx}
\usepackage{tablefootnote}
\usepackage{threeparttable}
\usepackage{multirow}
\usepackage{textcomp}
\usepackage{float}
\usepackage{breakurl}
\usepackage{autobreak}

\begin{document}
	
	
\title{Achieving ferroelectricity in a centrosymmetric high-performance semiconductor by strain engineering}
\maketitle


\author{Mengqi Wu}
\author{Zhefeng Lou}
\author{Chen-Min Dai}
\author{Tao Wang}
\author{Jiaqi Wang}
\author{Ziye Zhu}
\author{Zhuokai Xu}
\author{Tulai Sun}
\author{Wenbin Li*}
\author{Xiaorui Zheng*}
\author{Xiao Lin*}

\begin{affiliations}
	
Mengqi Wu, Chen-Min Dai, Jiaqi Wang, Ziye Zhu, Wenbin Li, Xiaorui Zheng\\
Key Laboratory of 3D Micro/nano Fabrication and Characterization of Zhejiang Province, School of Engineering, Westlake University, Hangzhou 310024, Zhejiang Province, P. R. China\\
Email: liwenbin@westlake.edu.cn; zhengxiaorui@westlake.edu.cn

\vspace{0.5em}
Zhefeng Lou, Tao Wang, Zhuokai Xu, Xiao Lin\\
Key Laboratory for Quantum Materials of Zhejiang Province, Department of Physics, School of Science, Westlake University, Hangzhou 310030, P. R. China\\
Email: linxiao@westlake.edu.cn

\vspace{0.5em}
Chen-Min Dai\\
Jiangsu Key Laboratory of Micro and Nano Heat Fluid Flow Technology and Energy Application, School of Physical Science and Technology, Suzhou University of Science and Technology, Suzhou 215009, China\\

\vspace{0.5em}
T. Sun\\
Center for Electron Microscopy, State Key Laboratory Breeding Base of Green Chemistry Synthesis Technology and College of Chemical Engineering, Zhejiang University of Technology, Hangzhou 310014, P. R. China\\
\end{affiliations}
~\\

\justifying  
	
\begin{abstract}
\noindent 
Phase engineering by strains in 2D semiconductors is of great importance for a variety of applications. Here, we present a study of strain induced ferroelectric (FE) transition on bismuth oxyselenide (Bi$_2$O$_2$Se) films, a high-performance (HP) semiconductor for next-generation electronics. Bi$_2$O$_2$Se is non-FE at ambient. Upon a loading force $\gtrsim 400$ nN, piezoelectric force responses exhibit butterfly loops on magnitude and 180$^\textrm{o}$ phase switching. By carefully ruling out extrinsic factors, these features are attributed to a transition to FE phase. The transition is further proved by the appearance of a sharp peak on optical second harmonic generation under an uniaxial strain. Fundamentally, solids with paraelectric at ambient and FE under strains are scarce. FE transition is discussed with the help of first-principle calculations and theoretical simulations. The switching of FE polarization acts as a knob for Schottky barrier engineering at contacts and serves as basis for a memristor with a huge switching ratio of 10$^6$. Our work endows a new degree of freedom to a HP electronic/optoelectronic semiconductor and the integration of FE and HP semiconductivity paving the way for multiple exciting functionalities, including HP neuromorphic computation and bulk piezophotovoltaic.
\end{abstract}
\setlength{\parindent}{1em}	
	
\section{Introduction}
\noindent Strain engineering is an effective way to tune physical properties and functionalities of tow-dimensional (2D)  materials~\cite{Kato2004Nature,Yang2020IM}. For instance, in NbOI$_2$, a surprisingly giant efficiency of optical second harmonic generation (SHG) is achieved by mechanical strains~\cite{Loh2022NP}. In non-centrosymmetric 3R-MoS$_2$, the bulk photovoltaic performance is boosted by stain induced polarization, dubbed bulk piezophotovoltaic~\cite{Iwasa2022NN}. In aforementioned examples, the strain plays a unique role in strengthening an existing property of a material. By contrast, strain tuning of phase transitions is of more fundamental interest, leading to profound modification of properties and phase engineering in 2D materials.

2D ferroelectricity (FE) has been observed in a number of layered materials of ultrathin film thickness with either in-plane~\cite{ChangK2016,ZhengCX2018SA,LinJH2022NC} or out-of-plane polarization ~\cite{Kalinin2015,LiuZ2016,LaiKJ2017,ZhangX2018PRL,Cobden2018Nature,Herrero2021Science,Gorbachev2022NN,Herrero2022NN,Lau2022Science}, such as monolayer SnTe~\cite{ChangK2016},  CuInP$_2$S$_6$~\cite{Kalinin2015,LiuZ2016}, multilayer WTe$_2$~\cite{Cobden2018Nature}, etc.  The discovery of 2D FE brought about a set of unprecedented physics~\cite{LiJ2021PNAS}, including interface sliding FE~\cite{LiJ2018JPCL,Shalom2021Science} and switchable ferroelectric metal~\cite{Cobden2018Nature}. Among 2D FE materials, of particular interest are FE semiconductors, which themselves may compose field-effect transistor (FET) channels, nevertheless with the capability of data storage, as expected in In$_2$Se$_3$~\cite{YePD2019NE,ZhangXx2021AM}. The memory in computation architecture, circumventing the bottleneck of von-Neumann frame, may help to break through Moore's limit upon Si-based industry.

Here, we demonstrate a strain engineering of FE transition in a high-performance (HP) semiconductor, Bi$_2$O$_2$Se (BOS) thin films. To our knowledge, strain induced FE transition in 2D layered semiconductors is yet to be explored. It was exclusive to quantum paraelectrics (QPE e.g. SrTiO$_3$ (STO)~\cite{Schlom2004Nature,Hwang2020NC} and KTaO$_3$~\cite{Uwe1975JPSJ}), which are perovskite insulators with 3D nature. BOS is a layered semiconductor, deemed as a highly competitive material for next-generation electronics~\cite{Wu2017} and optoelectronics~\cite{Khan2019,XuYB2019,LiuZ2019}, on account of environment stability, high mobility and robust band gap ($\triangle=0.8$ eV)~\cite{Wu2017}. The room temperature (room-$T$) Hall mobility $\mu_\textrm{H}$ is up to 450 - 500 cm$^2$.V$^{-1}$.s$^{-1}$~\cite{Wu2017} and low-$T$ $\mu_\textrm{H}$ is larger than 10$^5$ cm$^2$.V$^{-1}$.s$^{-1}$\cite{Peng2022NL}, surpassing that of MoS$_2$~\cite{Kis2011,Kis2013}. For BOS transistors, the on-state current density amounts to 1.3 mA $\mu$m$^{-1}$~\cite{Peng2022NL} with the leakage current below 0.015 A cm$^{-2}$, fulfilling the criteria of low-power HP electronic devices in the International Roadmap for Devices and Systems (IRDS)~\cite{Peng2022NE}.

Density functional theory (DFT) calculations predicted that upon moderate in-plane uniaxial/biaxial tensile strains, BOS evolves from body-centered tetragonal phase, space group I4/mmm  to FE orthorhombic phase, accompanied by the induction of giant in-plane lattice polarization 56.1 $\mu$C/cm$^2$~\cite{WuMH2017}. Parts of the authors predicted a giant modulation of electron mobility when approaching to the transition~\cite{LiWB2022}. Surprisingly, in BOS nanoplates, room-$T$ ferroelectricity was observed~\cite{Ghosh2019FE}, in stark contrast with the micro-scale thin films (non-FE), probably induced by lattice distortion.

Hereby, we present a study of FE transition on a micro-scale insulating BOS films by piezoelectric force microscopy (PFM) and optical second harmonic generation (SHG) measurements. Particularly, PFM was performed under a moderate loading force ($F_\textrm{L}$), that is distinct from a variety of literature. With $F_\textrm{L}\gtrsim 400$ nN, PFM signals evolve from trivial non-hysteretic curves to typical butterfly curves on amplitude and 180$^\textrm{o}$ switching on phase, which is compared with STO: a QPE and SiO$_2$: a common insulator. By carefully ruling out extrinsic contributions, the signals are attributed to intrinsic FE transitions. This is further corroborated by SHG measurements. SHG signals are negligible at ambient, while, exhibit a sharp peak upon uniaxial tensile strains. Combining theoretical simulations and DFT calculations, we unveil that FE emerges with in-plane polarization at $F_\textrm{L}\gtrsim 150$~nN. This is comparable with experimental observations. In the end, two-terminal memristive measurements are performed, which show a giant switching ratio about 10$^6$ (on-state current ($I_\textrm{on}$) about 1 $\mu$A and off-state current ($I_\textrm{off}$) about 1 pA). Our findings highlight a critical role played by strains in FE transitions of 2D materials. This phase engineering offers a venue for the study of a variety of emerging physics and appealing  functionalities.

\section{Results}

\subsection{Sample characterizations} 

\begin{figure*}[!thb]
	\includegraphics[width=15cm]{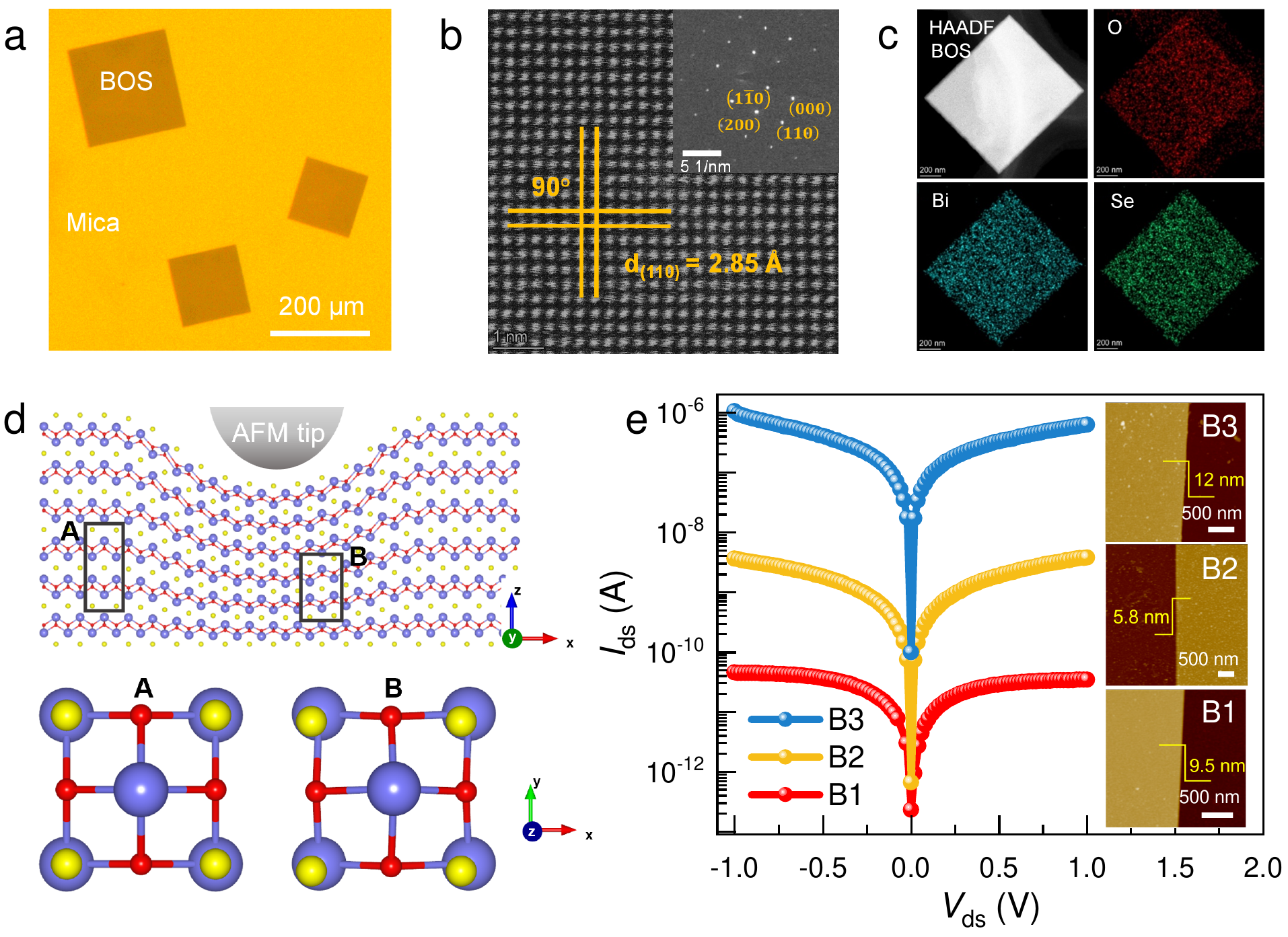}
	\caption{Sample characterization of insulating films. a) Optical image of BOS films on f-mica. b) Atomic-resolution HRSTEM image. The inset is the SAED pattern. c) High angle annular dark-field (HAADF) image in a large scale and the corresponding elemental mapping for Bi, O and Se. d) Upper panel: crystal structure of BOS thin films. The lattice is distorted under tip indentation. Low panel: top view of BOS unit cell. The lattice is centrosymmetric in undistorted region (A) and non-centrosymmetric in distorted region (B). e) I-V curves for three high resistance specimens. The insets are AFM topographic images, showing the thicknesses of three specimens.}
	\label{Fig1}
\end{figure*}

Figure~\ref{Fig1}a presents the optical image of square BOS films with a large size about $200\times200$~$\mu$m$^2$, grown on fluorophlogopite-mica (f-mica). The sharp atomic-resolution image along [001] zone axis from high-resolution scanning transmission electron microscopy (HRSTEM) is presented in Figure~\ref{Fig1}b. The inter-plane distance along [110] direction is about 0.285~nm, close to previous reports ~\cite{Wu2017,Wu2019nl}. The inset shows the selected area electron diffraction (SAED) pattern with sharp reflections.  In Figure~\ref{Fig1}c, the uniform distribution of Bi, Se, and O elements is verified by energy-dispersive X-ray spectroscopy (EDX) element mappings.

The crystal structure of BOS is shown in Figure~\ref{Fig1}d, in which [Se]$_n^{2n-}$ and [Bi$_2$O$_2$]$_n^{2n+}$ blocks stack alternately via weak electrostatic interactions~\cite{WangJL2020}. While under moderate nanoindentation, the lattice is distorted. The top layers experience inhomogeneous biaxial tensile strains, tangent to the tip surface. As a sequence, Se ions are expected to displace off-centering along [110] direction above a critical strain for paraelectric to FE transition~\cite{WuMH2017}. In common sense, the long range dipole interaction and switchable dipole orientation would be smeared by Thomas-Fermi screening from mobile electrons in highly-conducting samples as in previous reports~\cite{Wu2017,Khan2019}. Thus, we prepared three insulating specimens~\cite{WangT2022NR}, as seen in Figure~\ref{Fig1}e. All three specimens (B1, B2 and B3) exhibit non-Ohmic I-V curves with channel resistance altered by four orders of magnitude. The insets are AFM images indicating the thickness around 10 nm.

	~\\
\subsection{PFM characterizations under nanoindentation} 

\begin{figure*}[thb]
	\includegraphics[width=16cm]{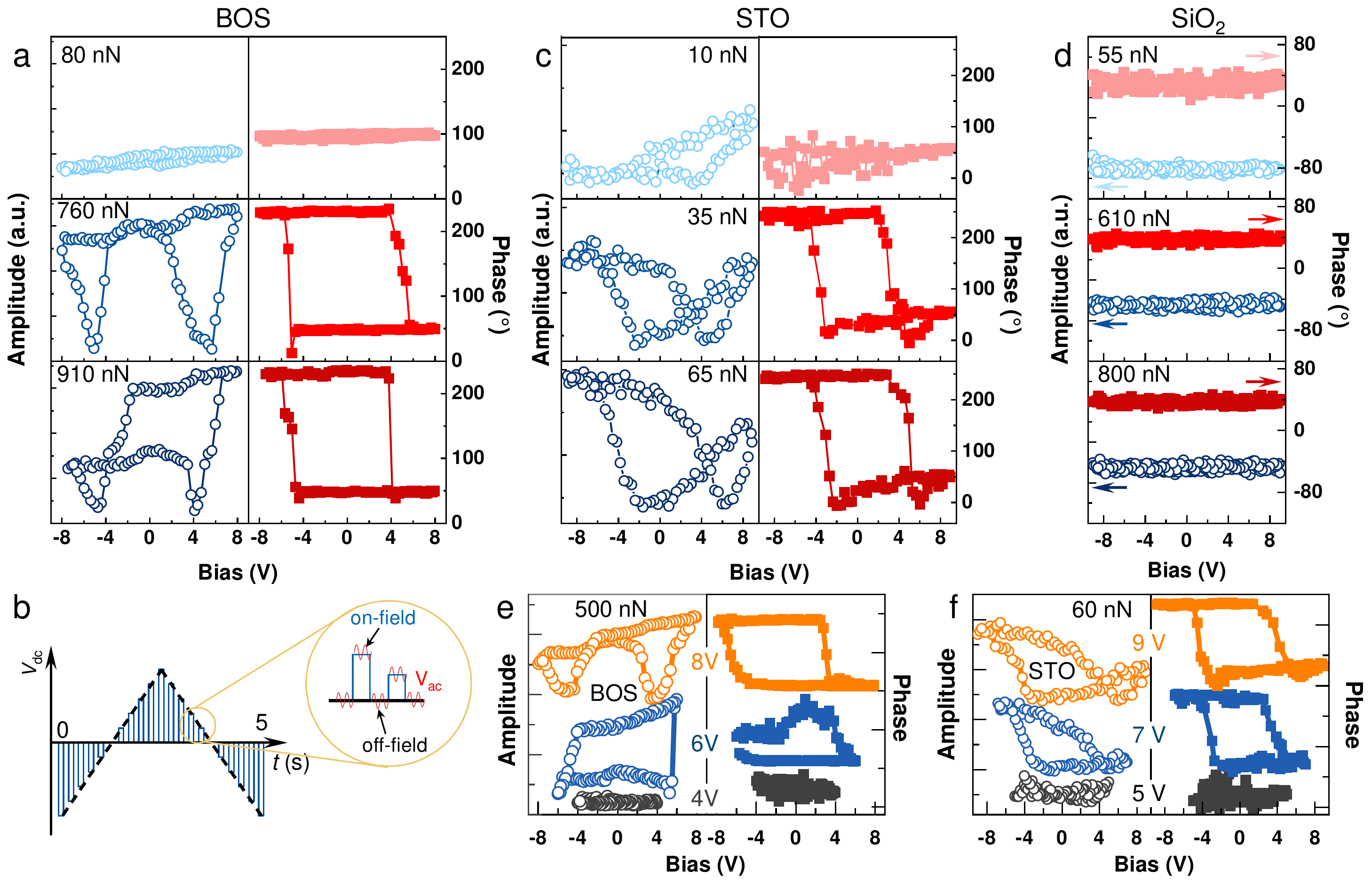}
	\caption{PFM measurements on off-field mode for BOS (B1) in comparison with STO and SiO$_2$. a) PFM signals for BOS under different tip forces. b) Triangular voltage waveform with frequency 0.2 Hz for PFM measurements. The off-field and on-field modes are measured at zero and finite dc voltage ($V_\textrm{dc}$), respectively. The ac voltage ($V_\textrm{ac}$) amounts to 0.8 V.  c,d) PFM signals for STO and SiO$_2$, respectively. e,f) The evolution of PFM response at various $V_\textrm{dc}$ for BOS and STO, respectively. The measurements were performed at a constant force in FE phase.}
	\label{Fig2}
\end{figure*}

Figure~\ref{Fig2} presents the off-field PFM measurements on the most insulating specimen B1, in comparison with STO bulk single crystals and 300 nm single-crystalline SiO$_2$ films. More PFM signals are presented in Figure~S1 and S2. As seen in Figure~\ref{Fig2}a, at ambient tip loading force ($F_\textrm{L}\approx70$~nN), both amplitude and phase are almost constant without apparent hysteresis when switching bias voltage between 8~V and -8~V. Under pressing with moderate force $F_\textrm{L}\gtrsim 700$ nN ($400$~nN in Figure~S1), a butterfly loop emerges, accompanied by 180$^\textrm{o}$ phase switching with a hysteresis width about 8V. Note that for the standard hysteresis measurement above, a triangular voltage waveform is applied, seen in Figure~\ref{Fig2}b. The off-field signal is measured at the interval between two pulsed biases.

In Figure~\ref{Fig2}c, similar signals, but at much smaller $F_\textrm{L}$, are observed in highly insulating STO (See more data in Figure~S3). This result is in accordance with the fact that STO inclined to be FE upon small tensile strains ($\varepsilon \lesssim 1\%$)~\cite{Schlom2004Nature,Hwang2020NC}. While, as a contrast, PFM signals remain flat and non-hysteretic through indentation on a common non-FE insulator SiO$_2$  in Figure~\ref{Fig2}d. It is intriguing to attribute the observation in BOS and STO to strain induced PE to FE transition. Though strain engineering of FE transition and polarization was predicted in several materials like Bi$_2$Se$_3$, Bi$_2$Te$_3$, etc.~\cite{XuH2020PRB,ZhengY2022JPCC}, to the best of our knowledge, such a transition, with PE at ambient and FE under strain, is scarce, exclusively proved in few systems, including QPE (e.g. STO~\cite{Schlom2004Nature,Hwang2020NC} and KTaO$_3$~\cite{Uwe1975JPSJ}).

\begin{figure*}[!thb]
	\includegraphics[width=12cm]{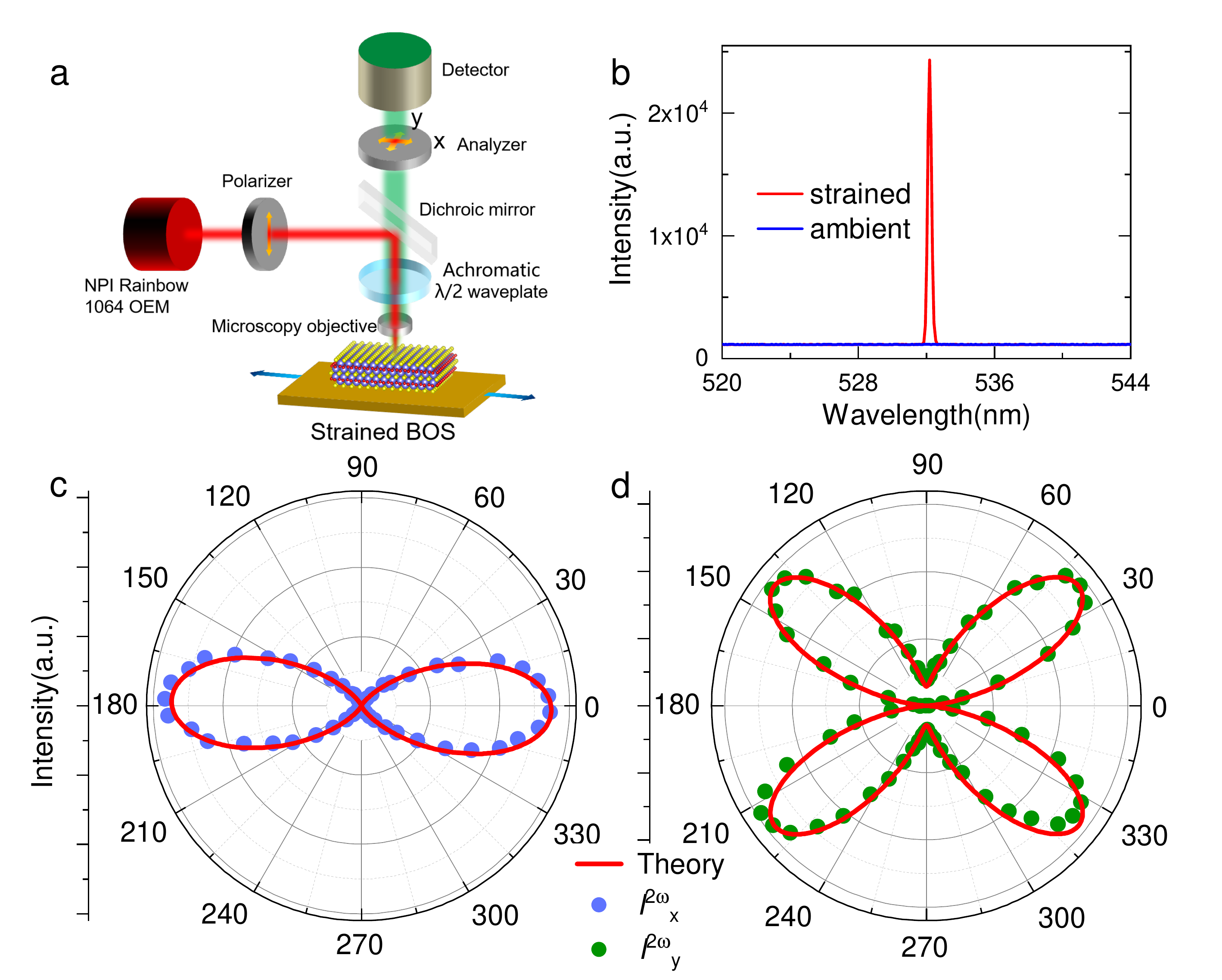}
	\caption{SHG measurements on BOS. a) Sketch of SHG apparatus. b) Unpolarized SHG signals at ambient and under moderate uniaxial tensile strain. The wavelength of incident light is 1064 nm.  c,d) Polarized SHG measurements under strain. $I^{2\omega}_\textrm{x}$ and $I^{2\omega}_\textrm{y}$ correspond to parallel and perpendicular modes, respectively. See more in Figure~S7. The solid lines are fits from mm2 point group~\cite{WuMH2017,LiWB2022}.}
	\label{Fig3}
\end{figure*}

Before pinning FE transition down, we should figure out that the phenomena on PFM are not due to several external factors. First, surface effect associated with absorbed molecules~\cite{Kim2012APL} and surface electrochemical reaction induced by water dissociation would cause hysteretic PFM signals~\cite{Dahan2006APL,Morozovska2007JAP}. However, in this case, PFM response should be similar at ambient and forced conditions. Second, current effect in leakage materials is another possible cause, which is nevertheless eliminated on off-field mode~\cite{Balke2015JAP}. We also note that Joule heating at ambient atmosphere led to hysteretic current switching in highly conducting BOS films~\cite{LiuBL2021AFM}. This process is irreversible, caused by oxidation, resulting in large modification of morphology. In contrast, our specimen is highly insulating and Joule effect is small. Moreover, the morphology is unchanged after PFM measurements (See Figure~S4). Third, ion motion is a significant cause for a fake PFM-like signal with a butterfly loop (even on off-field mode)~\cite{Chu2017AIP}. This process is slow with the relaxation time longer than the measuring time ($V_\textrm{dc}$ waveform 0.2 Hz). The distinction between ion motion and intrinsic FE is the following: the switching threshold voltage is field dependent in the former, while independent in the latter (coercive field - $V_\textrm{c}$). When $V_\textrm{dc}$ becomes smaller than $V_\textrm{c}$ in FE, the polarization switching is not allowed accompanied by the absence of PFM loops, which is exactly what we find in Figure~\ref{Fig2}e and f, in stark contrast with that from slow processes in which the loops remain~\cite{Chu2017AIP}. The ac voltage ($V_\textrm{ac}$) amplitude dependent signals are presented in Figure~S5. In the end, electrostatic interaction will significantly modify PFM responses, especially on on-field mode. While, on off-field mode, this modification is mild and will be discussed in detail below.

Putting Figure~\ref{Fig2}a and c under scrutiny, we find that butterfly loops become more asymmetric by further increasing $F_\textrm{L}$. More data is presented in Figure~S1-S3, according to which  the displacement of $V_\textrm{c}$ for STO is positive, while, for BOS, is nonmonotonous. This phenomenon is similar to what is caused by electrostatic effect~\cite{Kim2016APL,Kim2020APL}. Therefore, we present the expression of the first harmonic PFM response as the following:
	
\begin{equation}
	D_\textrm{ac}=d_\textrm{zz}V_\textrm{ac}+k^{-1}\frac{dC}{dz}(V_\textrm{dc}-V_\textrm{SP})V_\textrm{ac}
	\label{Eq1}
\end{equation}

\noindent $d_\textrm{zz}V_\textrm{ac}$ is the intrinsic term with $d_\textrm{zz}$ the piezoelectric coefficient. The second term is from the electrostatic effect, where $k$ is the contact stiffness of the cantilever, $C$ is the capacitance between tip and surface junction, and $V_\textrm{SP}$ the surface potential. Finite $V_\textrm{SP}$ is a conspicuous source of butterfly asymmetry on off-field mode.  $V_\textrm{SP}$ is sensitive to charge accumulation close to the surface, which, we argue, could be accounted by two possibilities: First, trapped charge injection during the sweeping; Second, intrinsically, the formation of flexoelectric polarization ($P_\textrm{FL}$) under strain gradient generated by $F_\textrm{L}$ (See more discussion below). To differentiate two terms, we performed multi-loop PFM sweeping at a constant $F_\textrm{L}$, as seen in Figure~S6. The loops of BOS present a negative displacement as loop number grows, while for STO, the loops are unshifted.  This highly implies that $P_\textrm{FL}$ dominates the asymmetry in STO, while in BOS, both terms are important. 
	
\begin{figure*}[!thb]
	\includegraphics[width=16cm]{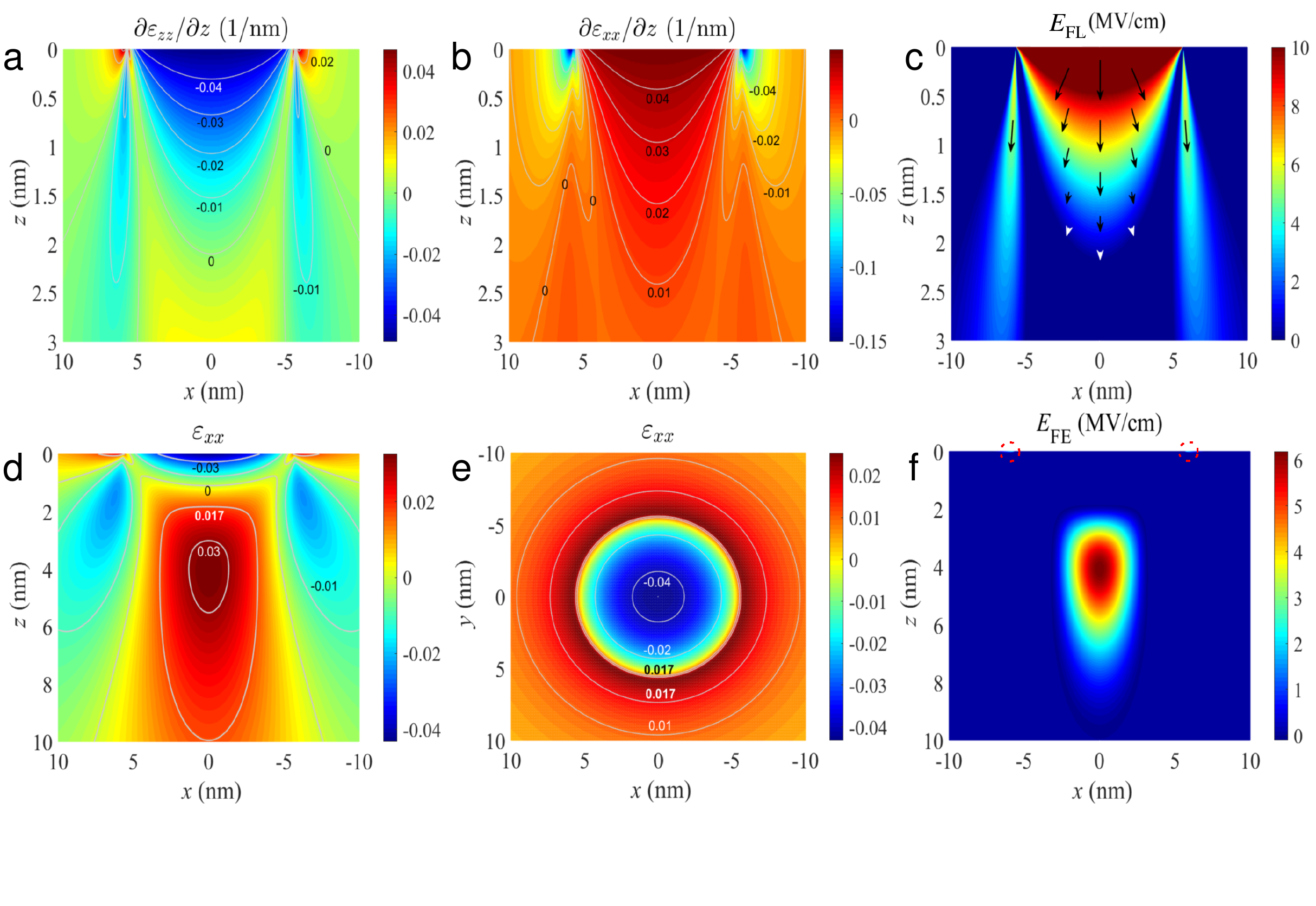}
	\setlength{\abovecaptionskip}{-30pt}
	\caption{Spatial distribution of strain gradients and strains at a tip loading force 800 nN obtained from theoretical simulation. a,b) Derivative of strains ($\varepsilon_\textrm{zz}$ and $\varepsilon_\textrm{xx}$) with respect to $z$. c) Spatial distribution of flexoelectric field ($E_\textrm{FL}$). d) Distribution of radial strain ($\varepsilon_\textrm{xx}$) in $zx$ plane. e) Distribution of radial strain at sample surface. f) Distribution of ferroelectric field ($E_\textrm{FE}$). The dashed circles mark weak FE region at surface. Polarization is along the radial direction. Note that $x$ and $y$ are in-plane axes; $z$ is the axis normal to surface.}
	\label{Fig4}
\end{figure*}

~\\
\subsection{Second harmonic generation}	

For more information, we performed SHG measurements on BOS films under uniform uniaxial strain in a reflection-mode SHG apparatus, seen in Figure~\ref{Fig3}a. In Figure~\ref{Fig3}b, no SHG peak is detected at ambient, in accordance with the centrosymmetry of lattice. By exerting moderate tensile strain, a sharp peak emerges, a manifestation of transition to noncentrosymmetric phase. Figure~\ref{Fig3}c and d show the angular dependence of polarized SHG signals for parallel-$I^{2\omega}_x$ and perpendicular-$I^{2\omega}_y$ modes respectively. The data is well fitted from $mm2$ point group by assuming the polar axis along the in-plane strain direction~\cite{Denev2011ProbingFU} (See more in Figure~S7 and Supporting Information.) The result demonstrates a breaking of C4 symmetry to C2 symmetry of orthogonal phase, as expected by DFT calculations~\cite{WuMH2017,LiWB2022}. The combination of PFM and SHG measurements and the distinct evolution of signals under strain provide a firm evidence for FE transition in BOS. 
	
~\\
\subsection{DFT calculations} 

\noindent To give a deep understanding of these phenomena, we performed theoretical simulations and present the spatial distribution of strain gradients and strains under nanoindentation in Figure~\ref{Fig4} (See Supporting Information). In Figure~\ref{Fig4}a and b, the strain gradient is as large as 10$^7$ m$^{-1}$ under a tip loading force $F_\textrm{L}=800$ nN, inducing apparent local inversion symmetry breaking in a centrosymmetric system, which is dubbed flexoelectricity. As seen in Figure~\ref{Fig4}c, such an effect generates a large electric field with a maximum 10 MV/cm within a narrow region close to the tip/sample interface. Its consequence, coined flexoelectronics, has become a focus of recent research~\cite{Das2019NC,WangZL2020NN}. 
	
For BOS, things are more intriguing, as seen in Figure~\ref{Fig4}d-f. In addition to flexoelectricity, we observe a bullet-like area right beneath the center of the contact region, corresponding to FE phase with radially distributed polarization. Moreover, a weak FE order occurs at surface close to the tip, marked by dashed circles in Figure~\ref{Fig4}f. According to DFT calculations, FE phase transition occurs in BOS with a in-plane biaxial tensile strain about 1.7\%, as seen in Figure~S8. In the tip-force model, this corresponds to 150 nN, above which FE phase emerges under nanoindentation. The calculation is in conformity with our observation that FE signals are absent at ambient and appear at $F_\textrm{L}\gtrsim 400$~nN. See more in Figure~S9 and Supporting Information.

\begin{figure*}[!thb]
	\includegraphics[width=18cm]{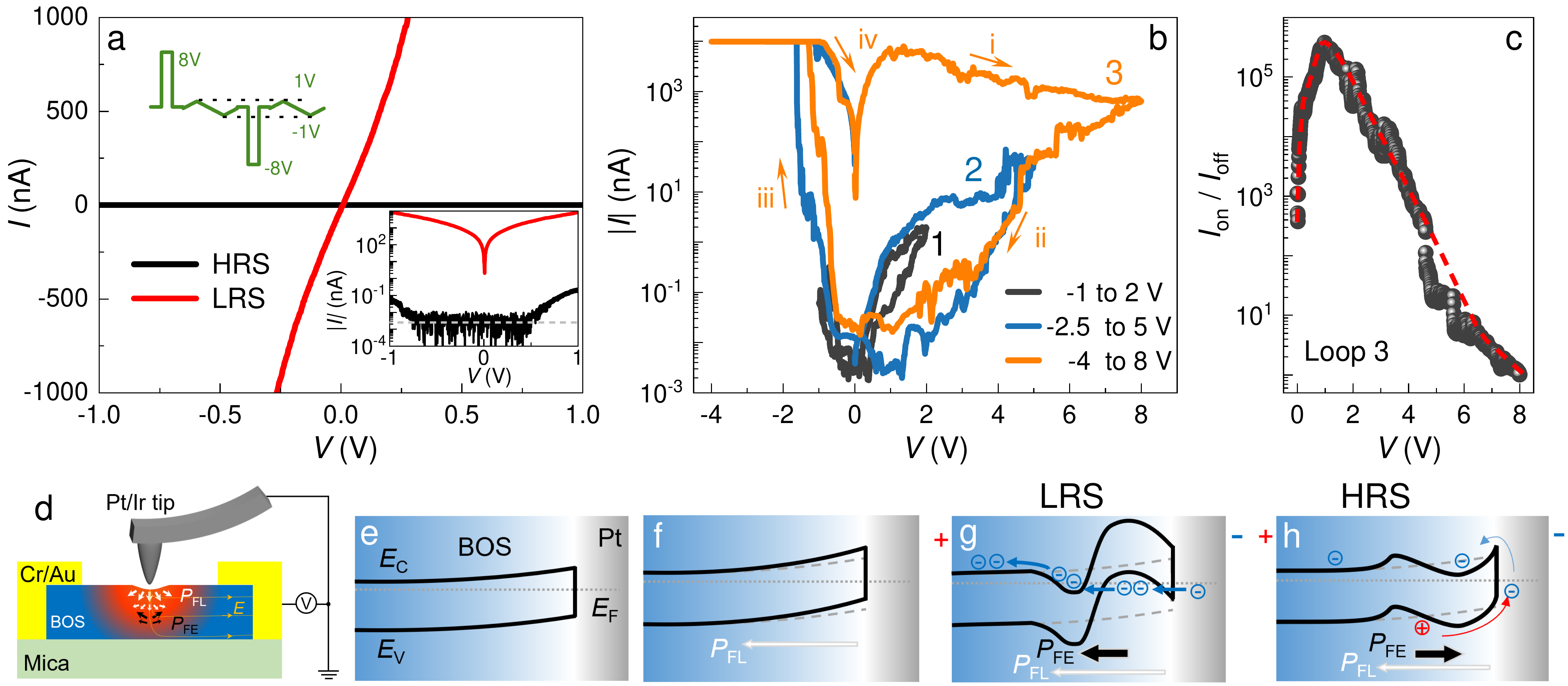}
	\caption{Memristive measurements on B3 at nanoindentation. a) I-V curves at low resistance state (LRS) and high resistance state (HRS), respectively. The upper inset presents the voltage waveform during the measurements. The lower inset is the semi-log plot. The resolution of the apparatus is at the level of pA. b) Current hysteresis loops. Referring to PFM measurements, the positive and negative maximum sweeping voltage are set asymmetrically. c) Switching ratio calculated from loop 3 in \textbf{b} at positive bias. d) Illustration of the set-up. e-h) Band alignment at tip/BOS interface under nanoindentation. $E_\textrm{c}$: conducting band; $E_\textrm{v}$: valence band; $E_\textrm{F}$: Fermi level; $P_\textrm{FL}$: FL polarization and $P_\textrm{FE}$: FE polarization.}
	\label{Fig5}
\end{figure*}
	
~\\
\subsection{Memristive measurements}

\noindent Having unveiled the fundamental ferroelectric properties, let us investigate the behaviors of ferroelectric memrisitive devices based on BOS. Figure~\ref{Fig5}a-c present the measurements of dc current ($I$) on B3 under moderate indentation. $I$ is substantially adjusted by poling the device prior to measurements. As seen in Figure~\ref{Fig5}a, positive poling (8 V) turns the device to a high resistance state (HRS) with negligible current of pA-level, while negative poling (-8 V) wakes up the device and $\mu$A level current is observed at the low resistance state (LRS). The inset is a semi-log plot and it is clearly resolved that the switching ratio between LRS and HRS ($I_\textrm{on}/I_\textrm{off}$ at 0.5 V) is extremely high, approaching to 10$^6$.  Note that $I_\textrm{off}$ is at the noise floor that sets the up bound of the signal, thus the real ratio can be higher if the instrument hosts better resolution. In spite of that, the switching ratio is already orders of magnitude higher than what was found in other ferroelectric semiconductors~\cite{ZhangXx2021AM,GuY2020AFM} and is among the highest value of sophisticated ferroelectric tunneling junctions based on ferroelectric insulators~\cite{WuD2020AM} (e.g. BaTiO$_3$~\cite{WuD2017NC,JinKJ2019iSci}).

In Figure~\ref{Fig5}b, the current hysteresis loops present a typical feature of memristors.  For the measurements, the dc bias ($V_\textrm{dc}$) sweeps from zero to $V_\textrm{max}$, then to $V_\textrm{-max}$ and back to zero. The hysteresis window gradually expands with increasing $V_\textrm{max}$, implying an enlarged switching ratio. In Figure~\ref{Fig5}c, $I_\textrm{on}/I_\textrm{off}$ ratio at sweeping peaks to $4\times10^5$ at $V\approx1$~V. We note that the current at HRS slightly levels up as the loop number grows, which might be caused by trapped charge injection that supplies a partial screening of local FE dipoles.
	
Below, we present a qualitative interpretation of the electric switching in terms of band modulation induced by mutual alignment of $P_\textrm{FL}$ and FE polarization ($P_\textrm{FE}$) . Figure~\ref{Fig5}d presents the illustration of the set-up: in which the side electrode was made of Cr/Au forming Ohmic contacts~\cite{WangT2022NR} and the ground is connected to Pt/Ir coated tip. As illustrated in Figure~\ref{Fig5}e, a Schottkey barrier forms at the interface between the tip and BOS, owing to the mismatch of work functions ($\varPhi_\textrm{Pr/Ir}>\varPhi_\textrm{BOS}$). In Figure~\ref{Fig5}f, $P_\textrm{FL}$ emerges towards BOS by slight indentation, bending up the bands. By further indentation into FE phase, $P_\textrm{FE}$ appears, providing barrier engineering through FE switching as shown in Figure~\ref{Fig5}g and h. Note that $P_\textrm{FE}$ is parallel to the surface according to calculations, which is hardly switched by the tip electric field in common cases. However, as seen in Figure~\ref{Fig5}d, the bending of lattice under indentation and the asymmetry of the contacts render more components of external electric field parallel to  $P_\textrm{FE}$ and allow for FE switching. The bands are mostly modified within the bullet-like area beneath the tip, since the FE phase at surface is rather weak. 
	

At LRS, $P_\textrm{FE}$ points opposite to the tip and introduces a kink-like structure to the bands. The conducting and valence bands cross the Fermi level, generating a Zener-like tunneling between each other. This effect leads to an enhanced conductance. When $P_\textrm{FE}$ is switched, a reversed 'kink' is introduced onto the bands, resulting in the absence of intersection between the Fermi level and the bands. Therefore, the tunneling conductance is vastly suppressed giving rise to HRS. 
	

~\\
\section{Conclusion}
	
\noindent In summary, we performed strain engineering of FE transition in a centrosymmetric HP semiconductor. Memristive measurements unveil a giant FE switching ratio of 10$^6$. For practical implications, FE adds a new dimension to an existing HP electronic and optoelectronic semiconductor. The built-in polarization expands the functionality of BOS to a broad range of exciting fields, of benefit to bulk piezophotovoltaics, nondestructive FE nonvolatile memories, strain-tuning optoelectronic logic devices, neuromorphic computation, etc. Fundamentally, adding FE degree of freedom to BOS renders this material a hunting ground of emerging physics, including spintronics, non-reciprocal transport, polar metal and even polar superconductivity.

~\\

\section{Experimental Section}

\textit{Sample fabrication and characterization:}
Highly-insulating BOS films were grown by chemical vapor deposition (CVD) in a two-zone furnace with  Se and Bi$_2$O$_3$ as precursors.  The optical image was obtained from ZEISS AXIO, ZOOM.V16 optical microscopy.  The aberration-corrected high-angle annular dark-field (HAADF) images were obtained by using a FEI Titan G2 80-200 ChemiSTEM operated at 200 keV, and energy dispersive X-ray spectroscopy (EDX) mapping was performed by using a Super-X EDX system with four silicon drift detectors for high sensitivity and capability. The electrical measurements were made on a Lakeshore CRX-4k probe station equipped with Keithley 4200A-SCS Parameter Analyzer.

\textit{AFM measurements:}
PFM were performed by using Asylum Research MFP-3D and conductive AFM (cAFM) were done on Asylum Research Cypher ES. Pr or Pr/Ir-coated tips with a radius $<$ 25 nm and a spring constant of $\sim$ 2 N m$^{-1}$ were utilized as probes. Each probe was calibrated by thermal calibration to obtain preciser spring constant and better inverse optical lever sensitivity (InvOLS). In contact modules (PFM and cAFM), the forces applied on tips are obtained by spring constant$\times$InvOLS$\times$$\Delta$deflection. 

For switching spectroscopy PFM measurements, the tip contacts sample surface at a small preset force. At the tip, dc bias ($V_\textrm{dc}$ square wave) steps over time, accompanied by a detecting ac voltage ($V_\textrm{ac}$ sin wave). For off-field mode, V$_\textrm{ac}$ is applied to detect piezoresponse signals at the interval between two steps ($V_\textrm{dc}=0$), as seen in Figure~\ref{Fig2}b, which effectively weakens the contribution of electrostatic interaction. In our measurements, PFM signals were detected at various tip forces. cAFM measurements were performed in ORCA module with a dual gain ORCA cantilever holder. The compliance current is 10 µA, and the noise floor is about 3 pA.


\textit{SHG measurements:}
SHG was performed in a confocal microscope (WITec, Alpha300RAS) under 1064 nm laser excitation (NPI Rainbow 1064 OEM). The angle dependent measurements for parallel-$I^{2\omega}_\textrm{x}$ and perpendicular-$I^{2\omega}_\textrm{y}$ geometry are operated by locking the mutual direction of incident and output polarizers at $0^\textrm{o}$ and $90^\textrm{o}$, respectively. For exerting uniaxial strain, the films were transferred to a flexible substrate (PI), then stretched by a homemade stretcher.  

\textit{First-principles calculations:}
The flexoelectric tensor is performed within the DFT local-density approximation (LDA)~\cite{Perdew1992PRB} using norm-conserving pseudopotentials as implemented in the ABINIT package~\cite{Gonze2009Abinit}. For the calculation of BOS, we use a tetragonal ten-atom unit cell of lattice constants ($a=b=7.3$ bohr, $c=22.86$ bohr) with a plane-wave cutoff of 60 Ha and a $7\times7\times2$ Monkhost-Pack mesh of $\textbf{k}$ points~\cite{Monkhorst1976PRB}. The unit cell is relaxed until atomic forces are smaller than $5\times10^{-5}$ Ha/bohr. The dielectric constant and Born effective charge ($Z^*$) are calculated by density functional perturbation theory. The ferroelectric polarization $P_\textrm{FE}$ was obtained from $P_\textrm{FE}=\sum Z_\textrm{i}^* D_\textrm{i}/V$, where the summation is over the atomic index i, $D$ is the displacement of atom, and $V$ is the volume of unit cell.

~\\
\noindent \textbf{Supporting Information}

\noindent Supporting information is available for this paper at the on-line version.

~\\

\noindent\textbf{Acknowledgments}

\noindent This research was supported by National Natural Science Foundation of China via Project 11904294 and 62004172. We thank the support provided by Lin Liu and Zhong Chen from Instrumentation and Service Center for Physical Sciences (ISCPS) and for Molecular Sciences (ISCMS) at Westlake University. We thank Westlake Center for Micro/Nano Fabrication for the facility support.

~\\
\noindent \textbf{Conflict of Interest}

\noindent The authors declare no conflict of interest.

~\\
\noindent\textbf{Author Contributions}

\noindent M.W., Z.L. and C.D. contributed equally to this work. M.W. performed AFM measurements supervised by X.Z.. Z.L. did SHG measurements and analyzed the data, supported by Z.X.. C.D. carried out first-principle calculations supervised by W.L. and supported by J.W. and Z.Z. T.W. grew BOS films. T.S. did STEM measurements.  M.W., Z.L., C.D. and X.L. prepared the figures. X.L. wrote the paper. W.L., X.Z. and X.L. led the project. All authors contributed to the discussion.

~\\

\noindent\textbf{Data Availability Statement}

\noindent The data that support the findings of this study are included in this article and its supporting information file and are available from the corresponding author upon reasonable request.

~\\
\noindent \textbf{Key Words}

\noindent Strain engineering, Ferroelectric transition, Bi$_2$O$_2$Se, Memristive 

~\\
\noindent \textbf{Notes}

\noindent We are aware that most recently, a paper (DOI: 10.1002/adma.202210854 ), appearing in Advance Materials, claimed the discovery of spontaneous out-of-plane polarization on as-grown Bi$_2$O$_2$Se films at ambient conditions. The result is in contradiction with our work and all the literature. After scrutinizing this paper, we believe that their data interpretation is misleading.


\bibliographystyle{MSP-1}
\bibliography{Maintext}

\begin{thebibliography}{10}
\providecommand{\url}[1]{\texttt{#1}}
\providecommand{\urlprefix}{URL }

\bibitem{Kato2004Nature}
Y.~Kato, R.~Myers, A.~Gossard, D.~Awschalom,
\newblock \emph{Nature} \textbf{2004}, \emph{427},  50.

\bibitem{Yang2020IM}
S.~Yang, Y.~Chen, C.~Jiang,
\newblock \emph{InfoMat} \textbf{2021}, \emph{3},  397.

\bibitem{Loh2022NP}
I.~Abdelwahab, B.~Tilmann, Y.~Wu, D.~Giovanni, I.~Verzhbitskiy, M.~Zhu,
  R.~Bert{\'e}, F.~Xuan, L.~d.~S. Menezes, G.~Eda, T.~C. Sum, S.~Y. Quek, S.~A.
  Maier, K.~P. Loh,
\newblock \emph{Nat. Photon.} \textbf{2022}, \emph{16},  644.

\bibitem{Iwasa2022NN}
Y.~Dong, M.-M. Yang, M.~Yoshii, S.~Matsuoka, S.~Kitamura, T.~Hasegawa,
  N.~Ogawa, T.~Morimoto, T.~Ideue, Y.~Iwasa,
\newblock \emph{Nat. Nanotechnol.} \textbf{2022}, ,  DOI: 10.1038/s41565.

\bibitem{ChangK2016}
K.~Chang, J.~Liu, H.~Lin, N.~Wang, K.~Zhao, A.~Zhang, F.~Jin, Y.~Zhong, X.~Hu,
  W.~Duan, Q.~Zhang, L.~Fu, Q.~Xue, X.~Chen, S.~Ji,
\newblock \emph{Science} \textbf{2016}, \emph{353},  274.

\bibitem{ZhengCX2018SA}
C.~Zheng, L.~Yu, L.~Zhu, J.~L. Collins, D.~Kim, Y.~Lou, C.~Xu, M.~Li, Z.~Wei,
  Y.~Zhang, M.~T. Edmonds, S.~Li, J.~Seidel, Y.~Zhu, J.~Z. Liu, M.~S. Tang,
  Wen-Xin adn~Fuhrer,
\newblock \emph{Sci. Adv.} \textbf{2018}, \emph{4},  eaar7720.

\bibitem{LinJH2022NC}
M.~Han, C.~Wang, K.~Niu, Q.~Yang, C.~Wang, X.~Zhang, J.~Dai, Y.~Wang, X.~Ma,
  J.~Wang, L.~Kang, W.~Ji, J.~Lin,
\newblock \emph{Nat. Commun.} \textbf{2022}, \emph{13},  5903.

\bibitem{Kalinin2015}
A.~Belianinov, Q.~He, A.~Dziaugys, P.~Maksymovych, E.~Eliseev, A.~Borisevich,
  A.~Morozovska, J.~Banys, Y.~Vysochanskii, S.~V. Kalinin,
\newblock \emph{Nano Lett.} \textbf{2015}, \emph{15},  3808.

\bibitem{LiuZ2016}
F.~Liu, L.~You, K.~L. Seyler, X.~Li, P.~Yu, J.~Lin, X.~Wang, J.~Zhou, H.~Wang,
  H.~He, S.~T. Pantelides, W.~Zhou, P.~Sharma, X.~Xu, P.~M. Ajayan, J.~Wang,
  Z.~Liu,
\newblock \emph{Nat. Commun.} \textbf{2016}, \emph{7},  12357.

\bibitem{LaiKJ2017}
Y.~Zhou, D.~Wu, Y.~Zhu, Y.~Cho, Q.~He, X.~Yang, K.~Herrera, Z.~Chu, Y.~Han,
  M.~C. Downer, H.~Peng, K.~Lai,
\newblock \emph{Nano Lett.} \textbf{2017}, \emph{17},  5508.

\bibitem{ZhangX2018PRL}
J.~Xiao, H.~Zhu, Y.~Wang, W.~Feng, Y.~Hu, A.~Dasgupta, Y.~Han, Y.~Wang, D.~A.
  Muller, L.~W. Martin, P.~Hu, X.~Zhang,
\newblock \emph{Phys. Rev. Lett.} \textbf{2018}, \emph{120},  227601.

\bibitem{Cobden2018Nature}
Z.~Fei, W.~Zhao, T.~A. Palomaki, B.~Sun, M.~K. Miller, Z.~Zhao, J.~Yan, X.~Xu,
  D.~H. Cobden,
\newblock \emph{Nature} \textbf{2018}, \emph{560},  336.

\bibitem{Herrero2021Science}
K.~Yasuda, X.~Wang, K.~Watanabe, T.~Taniguchi, P.~Jarillo-Herrero,
\newblock \emph{Science} \textbf{2021}, \emph{372},  1458.

\bibitem{Gorbachev2022NN}
A.~Weston, E.~G. Castanon, V.~Enaldiev, F.~Ferreira, S.~Bhattacharjee, S.~Xu,
  H.~Corte-Le{\'o}n, Z.~Wu, N.~Clark, A.~Summerfield, T.~Hashimoto, Y.~Gao,
  W.~Wendong, M.~Hamer, H.~Read, L.~Fumagalli, A.~V. Kretinin, S.~J. Haigh,
  O.~Kazakova, A.~K. Geim, V.~I. Fal’ko, R.~Gorbachev,
\newblock \emph{Nat. Nanotechnol.} \textbf{2022}, \emph{17},  390.

\bibitem{Herrero2022NN}
X.~Wang, K.~Yasuda, Y.~Zhang, S.~Liu, K.~Watanabe, T.~Taniguchi, J.~Hone,
  L.~Fu, P.~Jarillo-Herrero,
\newblock \emph{Nat. Nanotechnol.} \textbf{2022}, \emph{17},  367.

\bibitem{Lau2022Science}
L.~Rog{\'e}e, L.~Wang, Y.~Zhang, S.~Cai, P.~Wang, M.~Chhowalla, W.~Ji, S.~P.
  Lau,
\newblock \emph{Science} \textbf{2022}, \emph{376},  973.

\bibitem{LiJ2021PNAS}
M.~Wu, J.~Li,
\newblock \emph{Proc. Natl. Acad. Sci.} \textbf{2021}, \emph{118},
  e2115703118.

\bibitem{LiJ2018JPCL}
Q.~Yang, M.~Wu, J.~Li,
\newblock \emph{J. Phys. Chem. Lett.} \textbf{2018}, \emph{9},  7160.

\bibitem{Shalom2021Science}
M.~Vizner~Stern, Y.~Waschitz, W.~Cao, I.~Nevo, K.~Watanabe, T.~Taniguchi,
  E.~Sela, M.~Urbakh, O.~Hod, M.~Ben~Shalom,
\newblock \emph{Science} \textbf{2021}, \emph{372},  1462.

\bibitem{YePD2019NE}
M.~Si, A.~K. Saha, S.~Gao, G.~Qiu, J.~Qin, Y.~Duan, J.~Jian, C.~Niu, H.~Wang,
  W.~Wu, S.~K. Gupta, P.~D. Ye,
\newblock \emph{Nat. Electron.} \textbf{2019}, \emph{2},  580.

\bibitem{ZhangXx2021AM}
F.~Xue, X.~He, Z.~Wang, J.~R.~D. Retamal, Z.~Chai, L.~Jing, C.~Zhang, H.~Fang,
  Y.~Chai, T.~Jiang, W.~Zhang, H.~N. Alshareef, Z.~Ji, L.-J. Li, J.-H. He,
  X.~Zhang,
\newblock \emph{Adv. Mater.} \textbf{2021}, \emph{33},  2008709.

\bibitem{Schlom2004Nature}
J.~Haeni, P.~Irvin, W.~Chang, R.~Uecker, P.~Reiche, Y.~Li, S.~Choudhury,
  W.~Tian, M.~Hawley, B.~Craigo, A.~K. Tagantsev, X.~Q. Pan, S.~K. Streiffer,
  L.~Q. Chen, S.~W. Kirchoefer, J.~Levy, D.~G. Schlom,
\newblock \emph{Nature} \textbf{2004}, \emph{430},  758.

\bibitem{Hwang2020NC}
R.~Xu, J.~Huang, E.~S. Barnard, S.~S. Hong, P.~Singh, E.~K. Wong, T.~Jansen,
  V.~Harbola, J.~Xiao, B.~Y. Wang, S.~Crossley, D.~Lu, S.~Liu, H.~Y. Hwang,
\newblock \emph{Nat. Commun.} \textbf{2020}, \emph{11},  3141.

\bibitem{Uwe1975JPSJ}
H.~Uwe, T.~Sakudo,
\newblock \emph{J. Phys. Soc. Jpn.} \textbf{1975}, \emph{38},  183.

\bibitem{Wu2017}
J.~Wu, H.~Yuan, M.~Meng, C.~Chen, Y.~Sun, Z.~Chen, W.~Dang, C.~Tan, Y.~Liu,
  J.~Yin, Y.~Zhou, S.~Huang, H.~Q. Xu, Y.~Cui, H.~Y. Hwang, Z.~Liu, Y.~Chen,
  B.~Yan, H.~Peng,
\newblock \emph{Nat. Nanotechnol.} \textbf{2017}, \emph{12},  530.

\bibitem{Khan2019}
U.~Khan, Y.~Luo, L.~Tang, C.~Teng, J.~Liu, B.~Liu, H.-M. Cheng,
\newblock \emph{Adv. Funct. Mater.} \textbf{2019}, \emph{29},  1807979.

\bibitem{XuYB2019}
T.~Tong, Y.~Chen, S.~Qin, W.~Li, J.~Zhang, C.~Zhu, C.~Zhang, X.~Yuan, X.~Chen,
  Z.~Nie, X.~Wang, W.~Hu, F.~Wang, L.~Wenqing, P.~Wang, X.~Wang, R.~Zhang,
  Y.~Xu,
\newblock \emph{Adv. Funct. Mater.} \textbf{2019}, \emph{29},  1905806.

\bibitem{LiuZ2019}
Q.~Fu, C.~Zhu, X.~Zhao, X.~Wang, A.~Chaturvedi, C.~Zhu, X.~Wang, Q.~Zeng,
  J.~Zhou, F.~Liu, B.~K. Tay, H.~Zhang, S.~J. Pennycook, Z.~Liu,
\newblock \emph{Adv. Mater.} \textbf{2019}, \emph{31},  1804945.

\bibitem{Peng2022NL}
C.~Tan, J.~Jiang, J.~Wang, M.~Yu, T.~Tu, X.~Gao, J.~Tang, C.~Zhang, Y.~Zhang,
  X.~Zhou, L.~Zheng, Q.~Chenguang, P.~Hailin,
\newblock \emph{Nano Lett.} \textbf{2022}, \emph{22},  3770.

\bibitem{Kis2011}
B.~Radisavljevic, A.~Radenovic, J.~Brivio, V.~Giacometti, A.~Kis,
\newblock \emph{Nat. Nanotechnol.} \textbf{2011}, \emph{6},  147.

\bibitem{Kis2013}
B.~Radisavljevic, A.~Kis,
\newblock \emph{Nat. Mater.} \textbf{2013}, \emph{12},  815.

\bibitem{Peng2022NE}
Y.~Zhang, J.~Yu, R.~Zhu, M.~Wang, C.~Tan, T.~Tu, X.~Zhou, C.~Zhang, M.~Yu,
  X.~Gao, Y.~Wang, H.~Liu, P.~Gao, K.~Lai, H.~Peng,
\newblock \emph{Nat. Electron.} \textbf{2022}, ,  643.

\bibitem{WuMH2017}
M.~Wu, X.~C. Zeng,
\newblock \emph{Nano Lett.} \textbf{2017}, \emph{17},  6309.

\bibitem{LiWB2022}
Z.~Zhu, X.~Yao, S.~Zhao, X.~Lin, W.~Li,
\newblock \emph{J. Am. Chem. Soc.} \textbf{2022}, \emph{144},  4541.

\bibitem{Ghosh2019FE}
T.~Ghosh, M.~Samanta, A.~Vasdev, K.~Dolui, J.~Ghatak, T.~Das, G.~Sheet,
  K.~Biswas,
\newblock \emph{Nano Lett.} \textbf{2019}, \emph{19},  5703.

\bibitem{Wu2019nl}
J.~Wu, C.~Qiu, H.~Fu, S.~Chen, C.~Zhang, Z.~Dou, C.~Tan, T.~Tu, T.~Li,
  Y.~Zhang, Z.~Zhang, L.-M. Peng, P.~Gao, B.~Yan, H.~Peng,
\newblock \emph{Nano Lett.} \textbf{2019}, \emph{19},  197.

\bibitem{WangJL2020}
J.~Wang, J.~Wu, T.~Wang, Z.~Xu, J.~Wu, W.~Hu, Z.~Ren, S.~Liu, K.~Behnia,
  X.~Lin,
\newblock \emph{Nat. Commun.} \textbf{2020}, \emph{11},  3846.

\bibitem{WangT2022NR}
T.~Wang, Z.~Xu, Z.~Zhu, M.~Wu, Z.~Lou, J.~Wang, W.~Hu, X.~Yang, T.~Sun,
  X.~Zheng, W.~Li, X.~Lin,
\newblock \emph{Nano Research} \textbf{2022}, ,  DOI: 10.1007/s12274.

\bibitem{XuH2020PRB}
J.~Z. Zhao, L.~C. Chen, B.~Xu, B.~B. Zheng, J.~Fan, H.~Xu,
\newblock \emph{Phys. Rev. B} \textbf{2020}, \emph{101},  121407.

\bibitem{ZhengY2022JPCC}
Z.~Tang, M.~Dai, Y.~Chen, Q.~He, X.~Luo, Y.~Zheng,
\newblock \emph{J. Phys. Chem. C} \textbf{2022}, \emph{126},  10181.

\bibitem{Kim2012APL}
H.~Kim, S.~Hong, D.-W. Kim,
\newblock \emph{Appl. Phys. Lett.} \textbf{2012}, \emph{100},  022901.

\bibitem{Dahan2006APL}
D.~Dahan, M.~Molotskii, G.~Rosenman, Y.~Rosenwaks,
\newblock \emph{Appl. Phys. Lett.} \textbf{2006}, \emph{89},  152902.

\bibitem{Morozovska2007JAP}
A.~N. Morozovska, S.~V. Svechnikov, E.~A. Eliseev, S.~Jesse, B.~J. Rodriguez,
  S.~V. Kalinin,
\newblock \emph{J. Appl. Phys.} \textbf{2007}, \emph{102},  114108.

\bibitem{Balke2015JAP}
N.~Balke, S.~Jesse, Q.~Li, P.~Maksymovych, M.~Baris~Okatan, E.~Strelcov,
  A.~Tselev, S.~V. Kalinin,
\newblock \emph{J. Appl. Phys.} \textbf{2015}, \emph{118},  072013.

\bibitem{LiuBL2021AFM}
W.~Chen, R.~Zhang, R.~Zheng, B.~Liu,
\newblock \emph{Adv. Funct. Mater.} \textbf{2021}, \emph{31},  2105795.

\bibitem{Chu2017AIP}
Z.~Guan, Z.-Z. Jiang, B.-B. Tian, Y.-P. Zhu, P.-H. Xiang, N.~Zhong, C.-G. Duan,
  J.-H. Chu,
\newblock \emph{AIP Adv.} \textbf{2017}, \emph{7},  095116.

\bibitem{Kim2016APL}
B.~Kim, D.~Seol, S.~Lee, H.~N. Lee, Y.~Kim,
\newblock \emph{Appl. Phys. Lett.} \textbf{2016}, \emph{109},  102901.

\bibitem{Kim2020APL}
H.~Qiao, O.~Kwon, Y.~Kim,
\newblock \emph{Appl. Phys. Lett.} \textbf{2020}, \emph{116},  172901.

\bibitem{Denev2011ProbingFU}
S.~Denev, T.~T.~A. Lummen, E.~Barnes, A.~Kumar, V.~Gopalan,
\newblock \emph{J. Am. Ceram. Soc.} \textbf{2011}, \emph{94},  2699.

\bibitem{Das2019NC}
S.~Das, B.~Wang, T.~R. Paudel, S.~M. Park, E.~Y. Tsymbal, L.-Q. Chen, D.~Lee,
  T.~W. Noh,
\newblock \emph{Nat. Commun.} \textbf{2019}, \emph{10},  537.

\bibitem{WangZL2020NN}
L.~Wang, S.~Liu, X.~Feng, C.~Zhang, L.~Zhu, J.~Zhai, Y.~Qin, Z.~L. Wang,
\newblock \emph{Nat. Nanotech.} \textbf{2020}, \emph{15},  661.

\bibitem{GuY2020AFM}
M.~Gabel, Y.~Gu,
\newblock \emph{Adv. Funct. Mater.} \textbf{2021}, \emph{31},  2009999.

\bibitem{WuD2020AM}
Z.~Wen, D.~Wu,
\newblock \emph{Adv. Mater.} \textbf{2020}, \emph{32},  1904123.

\bibitem{WuD2017NC}
Z.~Xi, J.~Ruan, C.~Li, C.~Zheng, Z.~Wen, J.~Dai, A.~Li, D.~Wu,
\newblock \emph{Nat. Commun.} \textbf{2017}, \emph{8},  15217.

\bibitem{JinKJ2019iSci}
J.~Li, N.~Li, C.~Ge, H.~Huang, Y.~Sun, P.~Gao, M.~He, C.~Wang, G.~Yang, K.~Jin,
\newblock \emph{iScience} \textbf{2019}, \emph{16},  368.

\bibitem{Perdew1992PRB}
J.~P. Perdew, Y.~Wang,
\newblock \emph{Phys. Rev. B} \textbf{1992}, \emph{45},  13244.

\bibitem{Gonze2009Abinit}
X.~Gonze, B.~Amadon, P.-M. Anglade, J.-M. Beuken, F.~Bottin, P.~Boulanger,
  F.~Bruneval, D.~Caliste, R.~Caracas, M.~C{\^o}t{\'e}, T.~Deutsch,
  L.~Genovese, P.~Ghosez, M.~Giantomassi, S.~Goedecker, D.~R. Hamann,
  P.~Hermet, F.~Jollet, G.~Jomard, S.~Leroux, M.~Mancini, M.~J.~T. Oliveira,
  G.~Onida, Y.~Pouillon, T.~Rangel, G.-M. Rignanes, D.~Sangalli, R.~Shaltaf,
  M.~Torrent, M.~J. Verstraete, G.~Zera, J.~W. Zwanziger,
\newblock \emph{Comput. Phys. Commun.} \textbf{2009}, \emph{180},  2582.

\bibitem{Monkhorst1976PRB}
H.~J. Monkhorst, J.~D. Pack,
\newblock \emph{Phys. Rev. B} \textbf{1976}, \emph{13},  5188.

\end{thebibliography}

\clearpage


\end{document}